\documentclass[journal=jacsat,manuscript=article]{achemso}
\usepackage{amsmath} 
\usepackage{amssymb}
\usepackage{bm}
\usepackage{dcolumn}
\usepackage{graphicx}
\graphicspath{{.}{./figs/}} 
\usepackage{xcolor}
\usepackage{changes} 

\usepackage{footmisc}

\author{Abouzar Gharajeh\thefootnote{\ddag}}
\affiliation{Department of Electrical and Computer Engineering, The University of Texas at Dallas, Richardson, 75080, USA}

\author{Kiyanoush Goudarzi\thefootnote{\ddag}}
\affiliation{Department of Electrical and Computer Engineering, North Carolina State University, Raleigh, 27695, USA}
\alsoaffiliation{Current address:Department of Electrical Engineering, Pohang University of Science and Technology (POSTECH), 37673 Pohang, South Korea}

\author{Fargol Seifollahi}
\affiliation{Department of Physics and Astronomy, University of Texas Rio Grande Valley, Edinburg, 78539, USA}

\author{Qing Gu}
\email{qgu3@ncsu.edu}
\affiliation{Department of Electrical and Computer Engineering, North Carolina State University, Raleigh, 27695, USA}
\alsoaffiliation{Department of Physics, North Carolina State University, Raleigh, 27695, USA}

\title{Single-defect mode lasing in a non-Hermitian 1D trivial SSH lattice}

\begin{document}

%%%%% Equal authorship statement
\noindent\textbf{Equal Contribution:} Abouzar Gharajeh and Kiyanoush Goudarzi contributed equally to this work.

%%%%% date
\date{\today}

\begin{abstract}
We demonstrate robust single-mode lasing in a non-Hermitian 1D trivial Su-Schrieffer-Heeger defective lattice. 
This structure exhibits a significantly lower threshold power compared to the non-defective lattice, with single-mode lasing sustained across a wide range of pump powers.
Notably, our proposed device requires pumping of only the defect ring, leading to minimal required pump power.
Moreover, lasing is not confined to the edges but can occur at any point within the lattice. 
\end{abstract}

\maketitle
%\def\thefootnote{\ddag}\footnotetext{These authors contributed equally to this work.} 

%\renewcommand{\thefootnote}{\ddag}
%\footnotetext{These authors contributed equally to this work.}
%\footnotetext{These authors contributed equally to this work.}

\section{Introduction}

Non-Hermitian (NH) Hamiltonians introduce a new degree of freedom in physics by allowing for both real and complex eigenvalues, thus broadening the conventional framework of physical theories \cite{bender2007making}.
The introduction of NH Hamiltonians under parity-time (PT) symmetry, revealing real spectra, represents a groundbreaking advancement in quantum theories \cite{bender1998real}. 
However, despite its conceptual allure, the experimental confirmation of PT symmetry remains largely unexplored in condensed matter physics. 
This is primarily due to challenges such as the incoherence effect, the difficulty in introducing non-Hermiticity in a PT-symmetric manner, and the complexities arising from many-body interactions.
Fortunately, the experimental realization of PT symmetry can be pursued in the realm of photonics through the introduction of gain and loss media in an open system.
Consequently, NH photonics is an ideal platform for discovering novel functionalities including negative refraction \cite{fleury2014negative}, unidirectional invisibility \cite{lin2011unidirectional}, sensing \cite{mao2024exceptional}, while also offering a reliable solution for achieving single-mode lasing in on-chip photonic lasers \cite{11zhao2018topological,10parto2018edge,hodaei2014parity,st2017lasing,li2023topological}.

The emergence of topological physics has recently directed significant attention towards NH photonics by providing the control of wave propagation and confinement through the appearance of topological states that reveal robustness to large fabrication imperfections \cite{li2023topological, peng2024topological}. 
It offers anomalous relocation of the edge state \cite{6ramezani2022anomalous} and skin effect \cite{7yao2018edge,8martinez2018topological}, in which bulk states accumulate along the edge, forming localized bulk states, and has been used to funnel the light in optical lattices \cite{9weidemann2020topological}.
Topological lasing is another phenomenon in which by pumping the system, the edge state will be the first mode that reaches the lasing threshold \cite{10parto2018edge, ota2018topological, 11zhao2018topological}. 
More Specifically, by utilizing non-Hermiticity in a 1D array of microring resonators arranged to form a Su-Schrieffer-Heeger (SSH) model in a way that preserves PT symmetry in the system, one can have topologically protected lasing edge states \cite{10parto2018edge}. 
However, as the non-trivial topology of the structure dictates, these lasing states are limited only to the edges where the eigenmodes are localized. 
While one can possibly generate robust bulk states \cite{12ferdous2023observation}, such bulk states are not localized, thus might not be good candidates for lasing.

To overcome this limitation and achieve resilient localized states within a lattice, researchers have proposed a combination of two coupled non-Hermitian PT symmetric structures—one in the PT-exact phase and the other in the PT-broken phase \cite{13pan2018photonic}. 
This approach introduces a contrast between the imaginary components of the potentials in the two lattices, disrupting their coupling and resulting in lattices with nontrivial topology at their interface. 
Consequently, robust zero modes are generated within a 1D photonic lattice. 
Additionally, another study has employed a 1D non-Hermitian SSH structure containing a defect ring positioned between two trivial lattices \cite{11zhao2018topological}. 
This arrangement causes the defect ring to be isolated from the lattices, leading to a topological zero mode characterized by exponential decay. 
While these methods alleviate the constraint on topological lasing at the edge, there remains a requirement to strongly pump many sites to achieve the PT-broken phase and observe lasing emission effectively.

To overcome the high pump power requirement, a study previously proposed the utilization of degeneracies solely appearing in non-Hermitian systems---the so-called exceptional point (EP)---to create topologically protected defect states, despite the system being topologically trivial \cite{14mostafavi2020robust}. 
At the exceptional point, two or more eigenstates of the system coalesce and the Hamiltonian of the system becomes defective \cite{15miri2019exceptional,16ozdemir2019parity}. 
The existence of an EP allows us to induce a topological transition, even when the coupling configuration in the bulk—which completely determines the topological phase in Hermitian systems—does not change. 
Specifically, it is demonstrated that by means of a locally pumped defect embedded into the trivial phase of a laser array, localized symmetry-protected defect states can be created \cite{14mostafavi2020robust}.
These states which reside in the band gap are also manifested by the existence of an EP \cite{11zhao2018topological}.

In this letter, we present a theoretical analysis and experimental demonstration of a method to overcome the need for non-trivial topology and high pump power in topological lasers. 
Not only does our design not require lasing at the system's edges, we achieve robust single-mode lasing at desired points of the SSH microring array on demand. 
In addition, while previous studies relied on applying gain and loss uniformly across the sites of the SSH non-trivial array of microring resonators to induce lasing edge modes. 
Our approach does not pose such a restriction. 
We circumvent the necessity for high pump power by selectively pumping only one microring in the presence of a coupling defect within the array, allowing us to achieve robust single-mode lasing within the trivial SSH structure as desired.

%%%%% Fig.~ 1
\begin{figure}[t]
\includegraphics[width=4.5in]{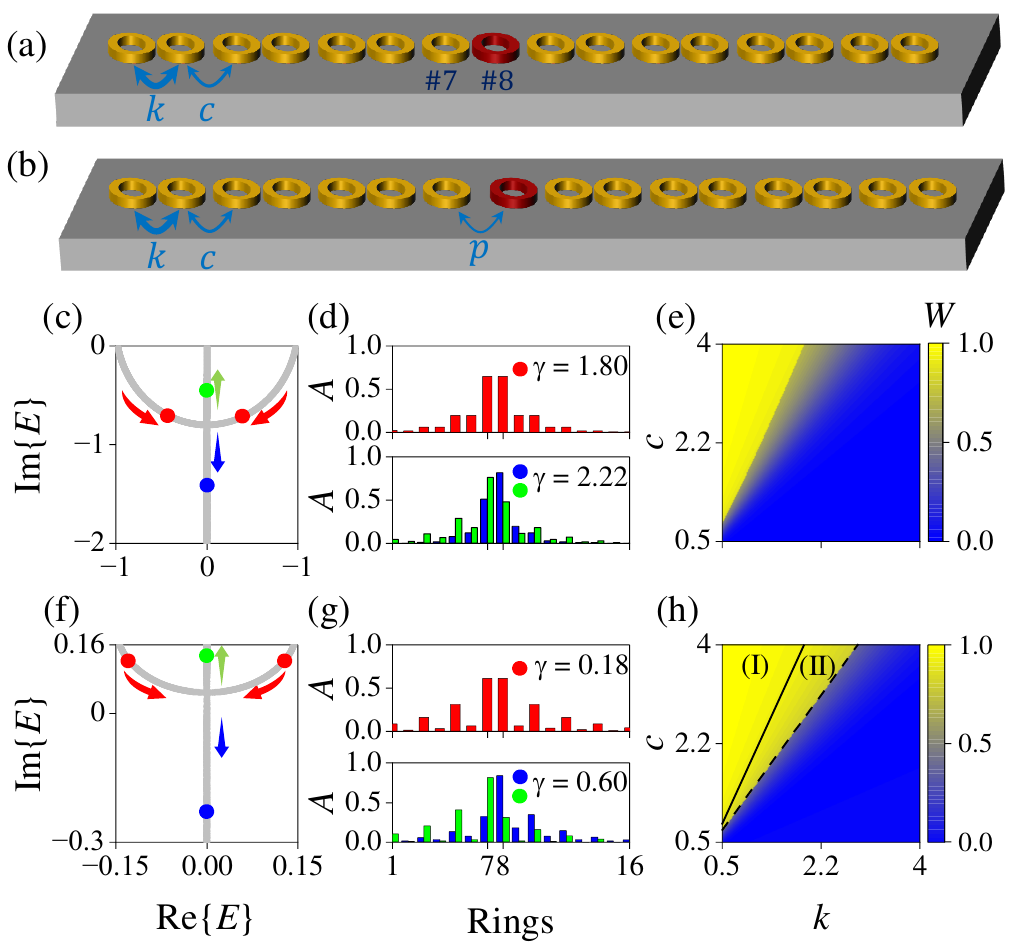}
	\caption{
		Schematic of (a) defectless and (b) defective, 1D SSH lattices. 
(c), (f) Eigenvalues of 7th and 8th rings for (a) and (b), respectively.
(d) Mode profiles for the defecless lattice at $\gamma = 1.8$ and $\gamma = 2.22$, respectively. 
(g) Mode profiles for the defective lattice at $\gamma = 0.18$ and $\gamma = 0.6$, respectively.
Winding number of (e) defectless  and (h) defective lattices.
$k = 1$ and $c= 0.5$ for (c), (d) and (f), (g). $p = 0.2$ and $\alpha = 0.2$ for (c)--(h).
In (a), (b) orange and red rings are loss and gain mediums, respectively.
In (c), the red dot represents $\gamma = 1.80$, while the blue and green dots represent $\gamma = 2.22$. In (f), the red dot corresponds to $\gamma = 0.18$, and the blue and green dots correspond to $\gamma = 0.60$. In both (c) and (f), arrows indicate the direction of increasing $\gamma$.
    	}
	\label{fig:fig1} 
\end{figure}

\section{Theory}

Fig.~\ref{fig:fig1}(a) depicts an open-boundary trivial SSH lattice with 16 sites, each represented by a microring, while Fig.~\ref{fig:fig1}(b) shows our proposed structure, with a defect. 
The orange and red rings are loss and gain media with onsite potentials of $i\alpha$ and $-i\gamma$, respectively.
The gain and loss rings refer to the pumped and unpumped rings, respectively, while $k$ and $c$ are intracell and intercell coupling coefficients, respectively.
The defective lattice is created by shifting the gain ring and subsequent rings to create a coupling $p$ between rings $\sharp7$ and $\sharp8$, as illustrated in Fig.~\ref{fig:fig1}(b).

We employed the tight-binding method to calculate the Hamiltonian of the structures, which are inherently non-Hermitian due to the presence of gain and loss rings (Supplementary Material).
Complex eigenvalues of both defectless and defective lattices indicate non-PT symmetric Hamiltonians, with EPs occurring at $\gamma_{E_1} = 2.00$ and $\gamma_{E_2}= 0.40$ as shown in Fig. S1 of the Supplementary Material.
It is evident that at the exceptional points, the eigenvalues coalesce, and due to the non-Hermitian nature of the Hamiltonian, the eigenvectors also become collinear.
% as depicted in Figs.~\ref{fig:fig1}(c) and (f), respectively. 
At $\gamma < \gamma_{E_1}$ and $\gamma < \gamma_{E_2}$ the lattices reveal bounded neutral oscillations as illustrated in the top panels of Fig.~\ref{fig:fig1}(d) and (g), respectively.
In the defectless case, by increasing $\gamma > \gamma_{E_1}$, ring $\sharp8$ radiates, and due to coupling, ring $\sharp7$ also radiates but with less amplitude than ring $\sharp 8$ as shown in Fig.~\ref{fig:fig1}(d) (bottom). 
This is confirmed by experimentally observed similar power distributions in rings $\sharp7$ and $\sharp8$ in the inset of Fig.~\ref{fig:fig4}(a).
In contrast to the defectless lattice, radiation in the defective lattice occurs at much lower pumping power ($\gamma_{E_2} << \gamma_{E_1}$). 
For $\gamma > \gamma_{E_2}$, the eigenvalues split into two complex branches with positive and negative values, as depicted in Fig.~\ref{fig:fig1}(f) representing attenuation and radiation for rings $\sharp7$ and $\sharp8$, respectively, as illustrated in the bottom panel of Fig.~\ref{fig:fig1}(g).
This is confirmed by experimentally observed localization of power in the ring $\sharp8$ as shown in the inset of Fig.~\ref{fig:fig4}(a).

In terms of topological nature, both lattices reveal a topological region labeled as (I) in Fig.~\ref{fig:fig1}(e) and (h).
Due to the introduction of a defect in the lattice, the defective lattice splits to two sublattices, in which the right one has the non-trivial nature. 
The appearance of the non-trivial sublattice creates a topological region in the defective lattice labeled as (II) in Fig.~\ref{fig:fig1}(h).
The yellow color indicates the topological region, while the blue color signifies the trivial region as shown in Fig.~\ref{fig:fig1}(e) and (h) \cite{song2019non}.
%In the proposed defective lattice, where $k>c$, the lattice exhibits trivial characteristics.  
As these lattices are finite and lack periodicity, the topological property is illustrated using the winding number ($W$) in real space for various values of $k$ and $c$: $W = 1$ denotes topological lattice, whereas $W = 0$ represents trivial lattice \cite{song2019non}.
The calculation method for the winding number in real space is summarized in Supplementary Material). 

%%%%% Fig.~ 2
\begin{figure}[!b]
\includegraphics[width=4.5in]{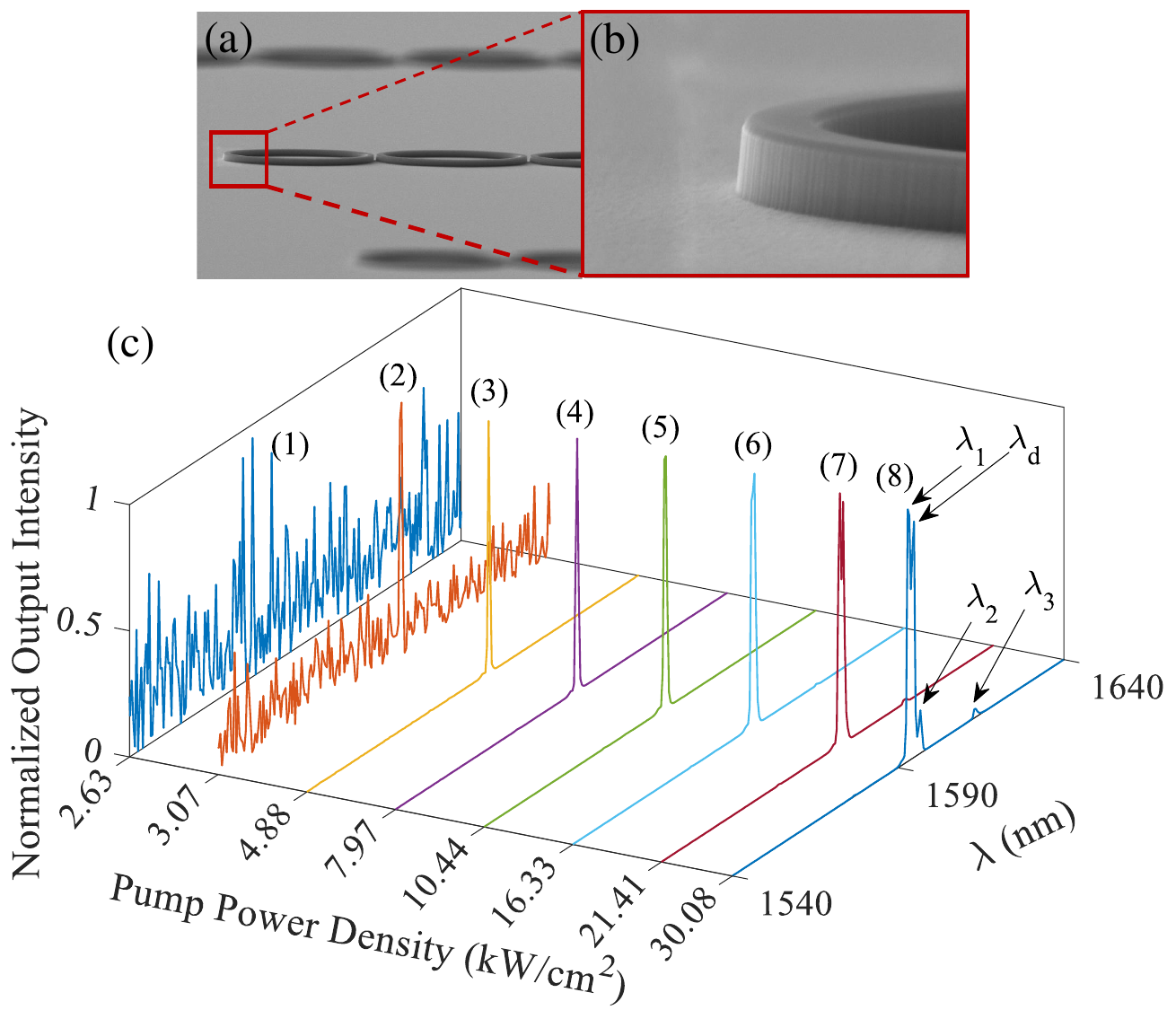}
	\caption{
	Scanning electron microscope images of fabricated (a) microring arrays and (b) a magnified section of a ring are presented. The images are captured before removing the HSQ e-beam resist.
(c) Evolution of emission spectrum with the pump power for the defective lattice.
	}
	\label{fig:fig2} 
\end{figure}

\section{Fabrication}

We fabricate the defectless and defective lattices.
The resulting lattices contain 16 identical microrings on a glass substrateas demonstrated in Fig.~\ref{fig:fig2}(a) and (b). 
The microrings consists of six quantum wells, each 10 nm thick, made of $\textrm{In}_{x = 0.564} \textrm{Ga}_{1-x} \textrm{As}_{0.933}\textrm{P}_{1-y}$, sandwiched between 20 nm thick barrier layers of $\textrm{In}_{x = 0.737} \textrm{Ga}_{1-x} \textrm{As}_{0.569} \textrm{P}_{1-y}$. 
The multilayer is capped by 10 nm thick layer of InP, resulting in a total thickness of 210 nm.
The radius, width, and hight of each rings are $5 \mu$m, 500 nm, and 210 nm, respectively.
The intracell and intercell distances are 150 nm and 200 nm, correspond to coupling strengths of 63~GHz and 38~GHz, respectively.
The distance defect between rings $\sharp 7$ and $\sharp 8$ is 900 nm correspond to a strength coupling of 415~MHz.
The details for fabrication steps are summarized in Supplementary Material.

\section{Experimental results}

To observe defect-mode lasing in the defective lattice, we pump only the defective ring with the laser spot size of $10 \mu$m, applying a circular shape Gaussian laser beam at $\lambda = 1064$ nm to the ring $\sharp 8$, defect ring.
At low pump powers, photoluminescence (PL) emission can be observed, marked as (1) and (2) in Fig.~\ref{fig:fig2}(c). 
%The PL emission has numerous frequency components and is not considered as lasing.
%In addition, the pumping power for the PL emission is bellow the threshold one, 3.5 $\textrm{kW/cm}^2$.
%As obvious, the less pump power is originated form applying defect as illustrated in Figs.~\ref{fig:fig1}(c) and (g).
By increasing the pump power above the threshold, 3.5 $\textrm{kW/cm}^2$, the defect ring initiates single-mode lasing at $\lambda_d = 1594.80$ nm, as indicated by (3)--(6) in Fig.~\ref{fig:fig2}(c).
Due to the minimal coupling between the defect ring and the adjacent rings of $\sharp 7$ and $\sharp 9$, the single-mode defect lasing persists over a wide range of pump power up to 16.33$\textrm{kW/cm}^2$, about five times the threshold pump power.
However, with further increase in the pump power, the coupling between the defect ring and ring $\sharp 7$ surpasses a critical value.
As a result, the degeneracy between their respective modes is broken, leading to the emergence of double-mode lasing at $\lambda_d = 1594.81$ nm and $\lambda_1 = 1592.89$ nm, as denoted by (7) in Fig.~\ref{fig:fig2}(b).
The splitting frequency resulted from broken degeneracy is proportional to the coupling coefficient between the rings \cite{hodaei2014parity}.
Upon further increasing the pump power up to 30 $\textrm{kW/cm}^2$, in addition to the coupling between the defect ring and ring $\sharp 7$, the coupling between ring $\sharp 9$ and the defect ring results in another broken degeneracy, consequently another lasing mode appears at $\lambda_2= 1596.72$ nm.
Furthermore, another competing mode of the defect ring acquires the sufficient modal gain, leading to lasing at $\lambda_3 = 1613.44$ nm.

%%%%% Fig.~ 3
\begin{figure}[!b]
\includegraphics[width=4.5in]{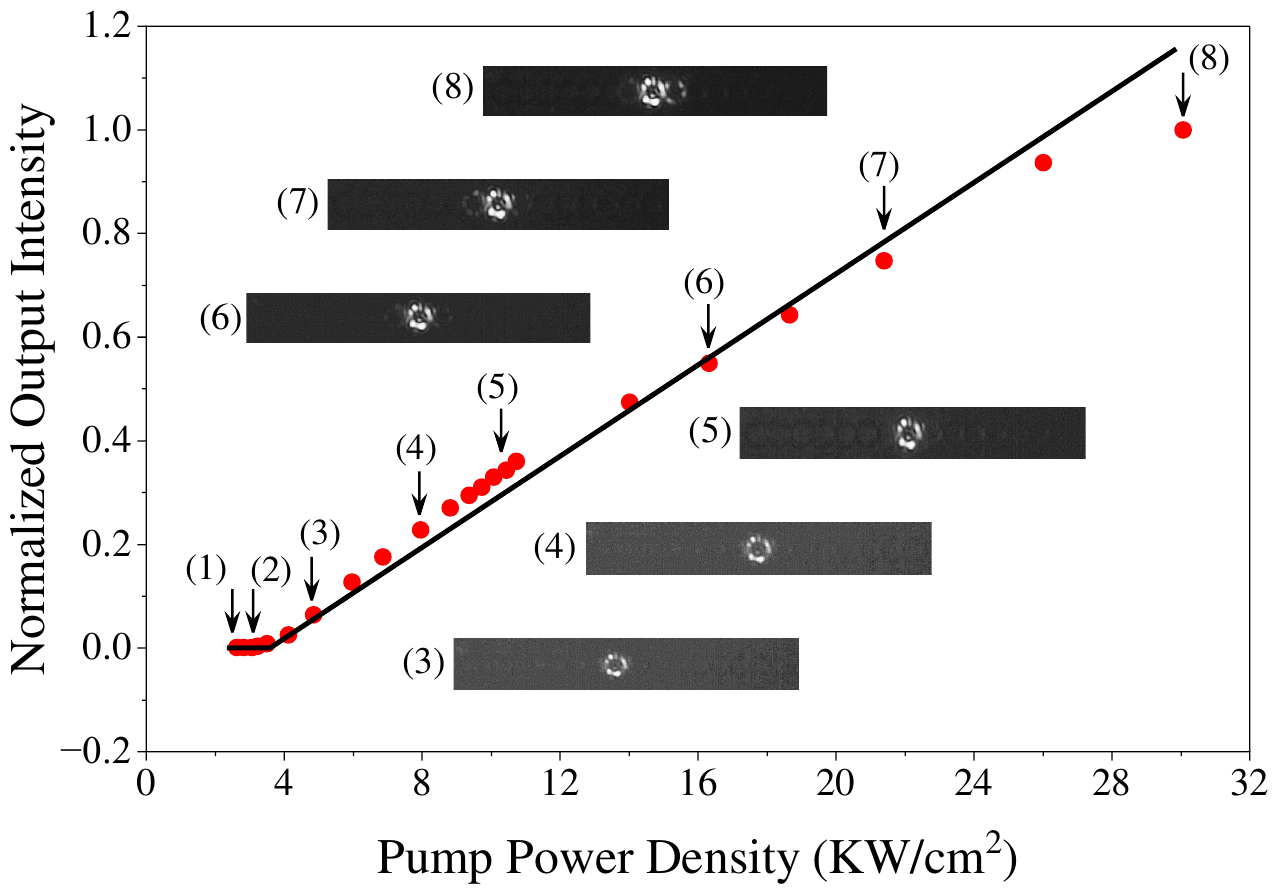}% Here is how to import EPS art
	\caption{
Characteristic light-light graph for the defective structure that shows spectral evolution at different pump power densities. 
The output intensities are measured at $\lambda_r = 1594.81$ nm.
The insets show the mode profile at different pump powers. 
The solid black line represents the fitted line for the measured data points represented by solid red circles.
}
	\label{fig:fig3} 
\end{figure}

%This part discusses the lasing characteristics and the respective mode profiles.
%The measurement of output intensities at $\lambda_r = 1594.80$ nm for different pump powers, depicted in Fig.~\ref{fig:fig3} as the light-light graph, reveals the characteristics of the defective laser.
%As discussed earlier, below the threshold power, lasing does not occurs, and the PL emission dominantes.
%However, with pump powers ranging from 3.5 to 16.33 $\textrm{kW/cm}^2$, the defect mode attains the sufficient gain, resulting in single-mode defect lasing with the lasing profile concentrated at the defect ring, as indicated by (3)--(6) in Fig.~\ref{fig:fig3}.
Focusing on the defective lasing mode at $\lambda_d = 1594.80$ nm, Fig.~\ref{fig:fig3} depicts its light-light curve as well as mode profiles, showing a lasing threshold of 3.5 $\textrm{kW/cm}^2$ for the defectdive mode, and a single-mode lasing range between 3.5 to 16.33 $\textrm{kW/cm}^2$ during which the light-light curve is nearly linear, and validated by the measured far-field emission profile that is concentated at the defect
ring (CCD camera images (3)--(6) in Fig.~\ref{fig:fig3}). Further increase in the pump power initiates multi-mode lasing, resulting in the power concentration in the defect ring and ring $\sharp 9$, marked as (7), followed by the power distribution across ring $\sharp 7$ and defect ring, and ring $\sharp 9$, marked as (8).

\begin{figure}[!t]
\includegraphics[width=4.5in]{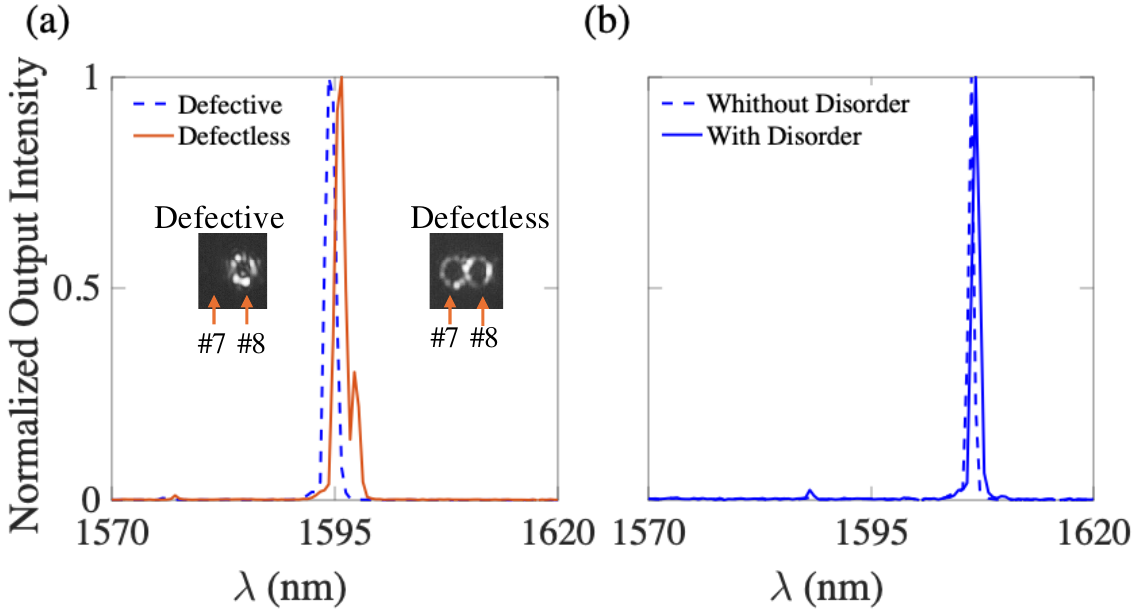}% Here is how to import EPS art
	\caption{
	Lasing spectrum from (a) a defective array in dashed blue color and a defectless array in solid red color. (a) The lasing spectrum for the defective array without random relocations (dashed blue color) and with random relocations (solid blue color). The mode profile of the defective and defectless lattices are shown in the inset of (a).
The pump powers for (a) and (b) are 14.021 $\textrm{kW/cm}^2$ and 7.64 $\textrm{kW/cm}^2$, respectively.
}
	\label{fig:fig4} 
\end{figure}

As we discussed earlier, introducing a defective structure with trivial topology leads to a decrease in the coupling between the defect ring and the adjacent rings, resulting in a drastic reduction in the threshold pump power compare to the trivial defectless lattice (Fig.~\ref{fig:fig1}(c) and (f)).
To demonstrate this reduction, we measure the lasing intensity at a pump power of 14.03 $\textrm{kW/cm}^2$ for both the defectless and defective structures. 
As depicted in Fig.~\ref{fig:fig4}(a), the output radiation intensity exhibits multi-mode lasing at $\lambda  =  1595.76$ nm and $\lambda = 1597.2$ nm for the defectless structure, while it shows a single-mode lasing at $\lambda = 1594.33$ nm for the defective structure. 
Consequently, the defective structure reveals lower threshold pump power than the defectless. 
%Owing to the single and two wavelength lasing of the defectless and defective structures, the corresponding mode profiles demonstrates the concentration of power in the defect ring, and defect ring and ring $\sharp9$ as illustrated in the insets of Fig.~\ref{fig:fig4}(a).
The far-field radiation pattern in the insets of Fig.~\ref{fig:fig4}(a) confirms the spectroscopic features.

Furthermore, we demonstrate the robustness of the defect-mode lasing by applying a position disorder along the array to the defective structure. 
The disorder is created by applying a random relocation of each ring up to 2 nm, which creates random couplings between the rings. 
Fig.~\ref{fig:fig4}(b) shows measured emission spectra at a pump power of 7.82 $\textrm{kW/cm}^2$ for defective structure with and without disorder.
In both cases, we observe single-mode lasing at $\lambda = 1606.79$ nm, with similar output intensities, which indicates the robustness of the defect mode. 
Note that the difference between lasing wavelengths in Fig.~\ref{fig:fig4}(a) and (b) is attributed to slight structural parameters such as the width and height of rings, due to variations in etching time.
To further check the robustness of the defect mode, we simulated the defective structure under the same position disordering, with random couplings (by adding random relocations to $a$ and $b$ distances (less than 5nm)). 
The calculation methods are discussed in Supplementary Material.

%%%%% Fig.~ 4

\section{Discussion}

We highlight four key innovations in this work. 
Firstly, we present, for the first time, experimental evidence of single-mode lasing from a non-Hermitian trivial 1D SSH defective lattice. 
While previous investigations have demonstrated single-mode topological lasers in 1D SSH and Aubry-André-Harper (AAH) topological lattices \cite{11zhao2018topological,hodaei2014parity,li2023topological}, our study explores a trivial defective lattice with reduced coupling, exhibiting a single defect lasing mode.
Secondly, our study demonstrates single-mode lasing over a large range of pump powers, up to 4.7 times the threshold power, surpassing previous 1D topological lasers. 
The decoupling of the defect ring from other rings in our configuration facilitates this large range of single-mode operation.
Thirdly, we reveal that introducing a defect into any ring facilitates single-mode lasing within the defect ring. This phenomenon arises because the introduction of the defect decouples the defect ring from the other rings, thereby enabling single-mode lasing to occur exclusively within it.
Lastly, pump power is specifically applied to the defect ring in our design, whereas previous studies require pumping of the entire structure. 
Looking into the prospect of the electrical pumping scheme, our design allows the electrical bias to the desired ring to achieve lasing from a specific location within our array.
%Moreover, our proposed structure eliminates the need for additional fabrication steps, such as depositing a Cr layer on certain rings to induce loss, which were required in previous approaches. 
%Furthermore, in our proposed structure, we experimentally demonstrate the robustness of the defect lasing mode to fabrication imperfections, including random relocations of the rings.
%These significant findings arise from the incorporation of a defective ring into our proposed lattice.

\section{Conclusion}

In conclusion, our study comprehensively investigates single-mode lasing within a trivial 1D SSH defective structure, both theoretically and experimentally. 
We establish the trivial nature of the defective structure through calculations of the winding number in real space. 
We experimentally reveal single-mode defective lasing at a low threshold of 3.5 $\textrm{kW/cm}^2$ as well as single-mode lasing across a large range of pump powers ranging from 3.5 to 16.33 $\textrm{kW/cm}^2$, at the defect ring. 
Importantly, as we demonstrated single-mode lasing at the defect ring, it is feasible to introduce a defect into any other ring, which decouples it from the rest of the structure and results in single-mode lasing within that ring.
Furthermore, our design exhibits robustness against random relocations of the rings, with a tolerance of up to 2 nm in the direction of the ring array. 
Moreover, instead of applying pump power to the entire structure, single-mode lasing can be achieved solely by pumping the defect ring, rendering our design highly efficient and amenable for electrical pumping. 
%Notably, our fabrication approach, which does not necessitate the use of a Cr layer on certain rings, proves cost-effective despite potential imperfections.
Overall, our work presents a novel class of non-Hermitian lasers characterized by high efficiency and versitility, rendering them well-suited for integration into photonic integrated circuits.


\end{document}